\documentclass[
]{ceurart}

\usepackage{listings}
\usepackage{svg}
\usepackage{adjustbox}
\usepackage{multirow}

\begin{document}

\copyrightyear{2025}
\copyrightclause{Copyright for this paper by its authors.
  Use permitted under Creative Commons License Attribution 4.0
  International (CC BY 4.0).}

\conference{Accepted in the IEEE Global Engineering Education Conference (EDUCON2025), London, UK, 22-25 April, 2025}

\title{LLM-Assisted Knowledge Graph Completion for Curriculum and Domain Modelling in Personalized Higher Education Recommendations}


\author[1]{Hasan Abu-Rasheed}[%
email=hasan.abu.rasheed@uni-siegen.de
]
\cormark[1]
\address[1]{Institute for Knowledge-Based Systems and Knowledge Management, University of Siegen, Siegen, Germany}
\address[2]{Institute for Embedded Systems, University of Siegen, Siegen, Germany}
\address[3]{Department of Digital Health Sciences and Biomedicine, University of Siegen, Siegen, Germany}

\author[2]{Constance Jumbo}[
]

\author[2]{Rashed Al Amin}[
]

\author[1,3]{Christian Weber}[
]
\author[2]{Veit Wiese}[
]

\author[2]{Roman Obermaisser}[%
]

\author[1]{Madjid Fathi}[%
]

\cortext[1]{Corresponding author.}

\begin{abstract}
  While learning personalization offers great potential for learners, modern practices in higher education require a deeper consideration of domain models and learning contexts, to develop effective personalization algorithms. This paper introduces an innovative approach to higher education curriculum modelling that utilizes large language models (LLMs) for knowledge graph (KG) completion, with the goal of creating personalized learning-path recommendations. Our research focuses on modelling university subjects and linking their topics to corresponding domain models, enabling the integration of learning modules from different faculties and institutions in the student’s learning path. Central to our approach is a collaborative process, where LLMs assist human experts in extracting high-quality, fine-grained topics from lecture materials. We develop a domain, curriculum, and user models for university modules and stakeholders. We implement this model to create the KG from two study modules: Embedded Systems and Development of Embedded Systems Using FPGA. The resulting KG structures the curriculum and links it to the domain models. We evaluate our approach through qualitative expert feedback and quantitative graph quality metrics. Domain experts validated the relevance and accuracy of the model, while the graph quality metrics measured the structural properties of our KG. Our results show that the LLM-assisted graph completion approach enhances the ability to connect related courses across disciplines to personalize the learning experience. Expert feedback also showed high acceptance of the proposed collaborative approach for concept extraction and classification.
\end{abstract}

\begin{keywords}
  Higher education \sep
  Knowledge graph (KG) \sep
  Large language models (LLM) \sep
  curriculum model\sep
  domain model \sep
  Recommender systems
\end{keywords}

\maketitle

\section{Introduction}

Higher education is a structured learning setting, where curriculums are designed to meet predefined learning goals for each study program. This leaves a small margin to meet the individual goals of different students in the same program, especially in programs where students have different study backgrounds, or come from various countries, such as international study programs. In these scenarios, the likelihood of a student repeating a course or a topic they already studied before becomes higher. In order to increase the efficiency of the full learning path that a student takes within their study, it is essential to consider their level of knowledge, the courses and modules they already studied, as well as their career and professional goals, to tailor their learning path to those individual needs, goals, and requirements.

However, the learning setting in higher education poses several challenges to accomplishing this personalization. On the one hand, teachers are occupied with many tasks \cite{bielikova_annotated_2020} and lack sufficient time to analyze the background knowledge of individual students, as well as to analyze the content of other courses in the university, which might be used to construct a personalized learning path. On the other hand, there is a lack of standard representation of the courses and study modules in different study programs, sometimes even inside the same university. To enhance the interoperability of different study programs, and thus the potential to create an efficient learning-path recommendation, we aim in this research to provide a foundation for creating a comprehensive and homogenic pool of learning materials, through a flexible and sufficiently-detailed representation of curriculums in higher education. 

We propose a human-AI collaboration approach to support teachers with intelligent, LLM-assisted algorithms that extract and analyze their educational materials. Automatic extraction is then evaluated and controlled by the human teacher to ensure its high-quality. To ensure the comparability of the learning materials amongst different study programs and different universities, we propose an ontological foundation that is designed specifically for higher education, covering the curriculum model, domain model, and student model. Our main research questions in this paper are therefore:
\begin{enumerate}
    \item How to develop a higher-education ecosystem, where LLM supports teachers in creating a knowledge graph for higher education?
    \item What are the sufficient classes of a higher education ontology, to integrate domain, curriculum, and user models?
\end{enumerate}

To answer those questions, we developed the collaboration process based on a KG, which serves as the network data structure that represents the three models and their educational content. The KG also forms the homogeneous database that allows a recommender system to find and recommend personalize learning paths.

\section{Background}
The recent developments of LLMs opened several doors to support teachers and experts in a faster extraction and analysis of the wide range of educational content, whether online or within universities and higher education institutions. In the following, we highlight some of the systems and approaches that have similar goals to our research.

\subsection{LLMs Support in Creating KGs}
LLMs are powerful tools for generating and enhancing KGs \cite{ibrahim_survey_2024}. They are designed to understand and generate natural language and are capable of a wide range of natural language processing tasks \cite{naveed_comprehensive_2023}. In the context of KGs, LLMs enhance KG embedding, completion, and construction, as well as supporting text generation and question-answering tasks \cite{pan_unifying_2023}. 
KGs are valuable means for representing the interconnected nature of information, especially in educational settings\cite{abu-rasheed_building_2023}. They serve as a solid data structure for organizing and storing knowledge \cite{kejriwal_knowledge_2022}. Kommineni’s study shows that integrating LLMs into KG creation can enhance entity recognition and relationship extraction streamlining KG creation\cite{kommineni_human_2024}. This is an essential tool in higher education where vast information must be organized cohesively. 

\subsection{LLM-Enhanced KG Creation for Higher Education}
Enhancing KG creation using LLMs in higher education introduces a robust method for modelling course content and structure \cite{sifaleras_educational_2024}. The traditional process of constructing KGs is largely dependent on human experts and this has proven to be more complex when constructing large-scale and high-quality KGs in domains that evolve constantly \cite{bielikova_annotated_2020}. 
Jhajj \cite{sifaleras_educational_2024} investigated the use of LLM in creating and enhancing educational knowledge graphs by analyzing three KGs. The human evaluators rated the graph created from LLM and contextual data (course syllabus) higher regarding its relevance and educational value. Together with course-specific information, LLM can extract information and generate nodes that have high semantic similarity with course materials enabling a more efficient semi-automated KG creation process in higher education.   
The fusion of KG and LLMs is synergistic, particularly in higher education. KGs can be used as factual context sources for LLM prompts to enhance learning recommendations \cite{abu-rasheed_knowledge_2024}. Evaluating the approach both qualitatively and quantitatively, results showed improvement in the quality of the generated text. Despite these use cases, LLM support for KG creation in higher education remains under-explored, particularly in comprehensive multi-dimensional models, linking users, curriculums, and domain elements. 

\subsection{A Comprehensive Ontology for Higher Education }
Ontologies are formal representations of data, which in our use case provide a holistic view of higher education requirements, incorporating curriculum, user, and domain models. Recently, ontologies have been commonly used in higher education \cite{stancin_ontologies_2020}. However, traditional ontologies focus on user and educational models \cite{gangemi_developing_2018}, which represent the subject matter within the educational environment. This underscores the need for providing a detailed and comprehensive representation of relevant knowledge and concepts in real-world domains. Including a domain model provides contextual depth, which is important for recommending courses that align with user backgrounds, goals, and career orientations. LLM-support for curriculum and domain modelling offers a solid foundation for personalized learning recommendations in higher education. 

\section{Methodology}
To accomplish creating a graph data structure for modelling content in higher education and supporting teachers to represent their existing materials in this model, we propose a four-step implementation strategy: Graph ontology definition, automated topic extraction, human-AI collaboration for validation task, and KG construction.

\subsection{Defining a Higher Education Ontology}
\begin{figure}
  \centering
  \includegraphics[width=0.9\linewidth]{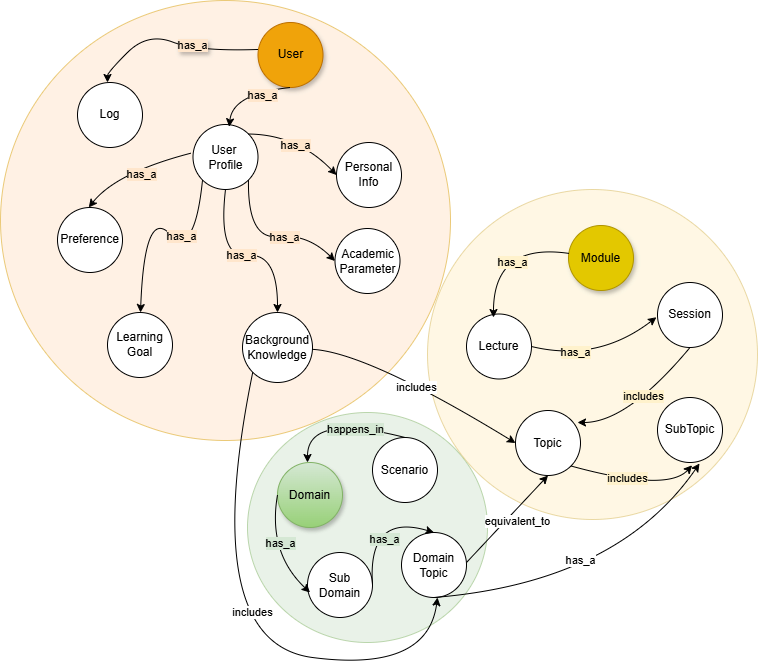}
  \caption{Proposed ontological structure of the knowledge graph, with domain, curriculum, and user models.}
  \label{Fig:1}
\end{figure}

 To develop a graph structure that accommodates learning content from different study programs and universities, we developed the content representation models as interoperable ones. This meant selecting ontological classes that apply to a wider range of course structures, enabling teachers to adapt their learning content to that model easily. 
To that end, we select the class “Topic” as the core taxonomical element that is used to break down the content of each session within the study module, see Figure \ref{Fig:1} As topics in the learning content differ in their level of detail, we add a “Sub-Topic” class to account for more granularity in the educational content. In this sense, a Topic is defined as an abstract concept, which is being taught in the learning session, while a Sub-Topic is defined as the fine-grained content, which is described through a tangible means, such as lecture slides. A Session is then the iterative instance of a Lecture, which takes place, e.g., on a weekly basis.
The proposed ontology is composed of three models:
\begin{enumerate}
    \item The curriculum model: which includes the classes representing educational content as defined by the university.
    \item The domain model: which is used to model a higher abstraction level of the domain of knowledge. The domain model includes domains to which the learning content belongs, such as Automotive, Embedded Systems, or Communication domains. In each domain, more granular concepts are represented as Sub-Domains, such as Self-Driving Cars being a sub-domain of the Automotive domain.
    \item The user model: which is meant to represent the stakeholders in the educational setting, focusing on students and the factors that can be used for learning personalization. These include the background knowledge, learning goals, preferences, as well as academic parameters, as adopted from \cite{hotho_educor_2021}, which represent all other academic aspects related to the student’s learning process, such as their exam performance.
\end{enumerate}

\subsection{Automatic Content Extraction and Classification}
In order for teachers to classify the content of their lectures into topics and sub-topics, a great amount of time and effort needs to be invested, which is rarely available for teachers due to their busy schedules. Therefore, we developed an automated extraction and classification pipeline, to support teachers with this task and considerably reduce the time and effort requirements on their side. 

\begin{figure}
  \centering
  \includegraphics[width=0.5\linewidth]{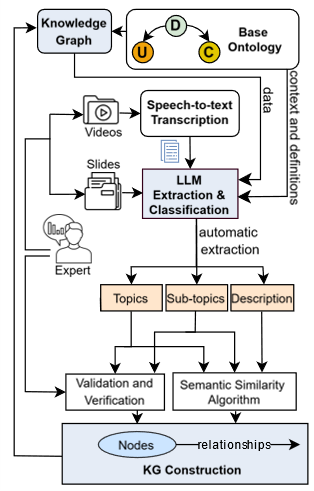}
  \caption{Pipeline components for the transcription, extraction, classification, and KG construction, based on a human-AI collaborative approach.}
  \label{Fig:2}
\end{figure}

The pipeline, see Figure \ref{Fig:2}, begins with the extraction of textual content from the learning materials, including lecture slides, manuscripts, and lecture video recordings. While the extraction of the textual data from lecture slides and manuscripts is relatively straightforward, the extraction of the text from lecture videos requires transcribing these videos. Here, we assume that the figures, images, and equations in the slides are annotated or described textually. We conduct this transcription automatically utilizing the advanced capabilities of OpenAI’s Whisper model \cite{radford_robust_2022}, which is a speech recognition model with high accuracy in zero-shot recognition tasks. Video transcripts are then fed, alongside the other textual content materials to an LLM for the topic extraction and classification tasks.

In the topic extraction step, we utilize OpenAI’s GPT-4o model. The LLM is prompted to extract the main concepts that each session includes. A concept represents here a key topic that is being taught in the session. Extracted concepts are then classified into topics and sub-topics based on the definitions of these two classes in our ontology. A topic is defined in the LLM prompt as an abstract concept, which is taught within the session. A sub-topic is defined as the concept that is being explained, in detail, through learning materials (e.g. lecture slides). The LLM is also instructed to consider the hierarchical structure of the two classes, where each topic includes one or more sub-topics. A topic is then an overarching concept, whose sub-topics are the actual content of the lecture. It is important to point out here that this ontology-based contextualization of the LLM prompt is imperative to extract an accurate representation of the learning content, ensure a sufficient level of interoperability, and enable finding meaningful similarities between the topics and sub-topics among lectures from different programs or universities.

In addition to the topics and sub-topics, the detailed content of lecture slides and videos provides a solid source for generating high-quality descriptions, especially on the sub-topic level, since it is the concept explained in detail in the session. We use those descriptions, in addition to the titles of the topics and sub-topics, to search for semantic similarities between different study programs.

\subsection{Human Validation of the Automated Extraction}
The risks of LLMs in automatic extraction, classification, and generation tasks, are high in a sensitive field like education. Therefore, we ensure the high-quality of topic extraction and classification through a human-in-the-loop approach, in which teachers validate the predictions of the intelligent algorithm, and make the final decision on extracted concepts and their classification as topics or sub-topics.
The preliminary extraction and classification by the LLM serves here as an advanced starting point for the teachers and human evaluators, which saves them a considerable amount of time when analyzing the educational content. This also applies to the automated descriptions of topics and sub-topics, since the phrasing of the description is based on the session content itself and not the general knowledge of the model, i.e., in its training data. The final decision on the topics, sub-topics, and their descriptions is made by the teachers themselves, to ensure a high-quality content curation of the KG, which is necessary to enable generating high-quality learning-path recommendations for the students.
The role of the teachers in our system is not limited to validating the last prediction of the intelligent models. Teachers are also integrated in the contextualization of the LLM prompts by providing input on the definitions, limitations, and rules that the model should consider when analyzing, extracting, and classifying the topics of their learning materials. An example of those definitions is the specific meaning of terminology that is used in the lecture, which may differ, slightly or considerably, among different domains.

\subsection{Knowledge Graph Construction}
Based on the validated content from the teachers, the KG construction process is then conducted through two steps:
\begin{enumerate}
    \item Creating graph nodes: according to the base ontology, and populating them with the data from the learning content, domain, and users.
    \item Creating graph relations: which is conducted on two levels: the hierarchical structure from the base ontology, and the semantic similarity between the topics and sub-topics of different lectures.
\end{enumerate}
 To calculate the semantic similarities, we first build on the natural language processing pipeline in [5], where the title and description properties of the learning content are utilized to find similar content. This approach fits our extraction process of the topics, sub-topics, and their descriptions. Secondly, we utilize the LLM itself to analyze the extracted content and search for similar content from other lectures.
 
We use a semi-supervised approach for the validation of semantic similarities, where the human teacher validates a random sample of the predicted relations in the KG. This approach is adopted due to the large amount of semantic relations that can be identified when the KG grows to include large-scale educational content. Through teacher validation, hints are collected to adjust the parameters and prompts of the relation extraction pipeline and the LLM, respectively, sustaining the collaborative approach in our overall solution.

Once the KG is constructed, it also becomes a source of data that is fed to the LLM, to provide contextual information for the extraction and classification tasks. We found this method particularly effective when the topic extraction and classification are done to a lecture that is contextually similar to the existing content of the KG, such as topics from the same domain, or the same field of application.

\section{Evaluation and Results}

To evaluate the proposed pipeline, we construct the knowledge graph from the content of two university modules, namely Embedded Systems and Development of Embedded Systems Using FPGA. The two courses are taught in the University of Siegen, Germany, for master- and bachelor-level students.  These courses were chosen because they are taught separately, despite the potential overlap between them. Each course is composed of 9 sessions, spanning the lecture period of one academic semester. Each weekly session covers a diverse number of topics, which differ in their level of detail.

\subsection{Expert Evaluation of the Ontology and Pipeline}
We conduct the evaluation with module lecturers, to ensure high-quality feedback based on their experience with the module’s content and learning goals. Evaluators included one professor and two PhD assistants, working in the Embedded Systems Institute in the same university. We collect qualitative feedback from the lecturers about the base ontology, and the overall acceptance of our approach for human-AI collaboration. Feedback is collected through group meetings with the lecturers, in which the ontology was presented and evaluated from the perspective of use cases they face in their teaching. 

Moreover, we collect quantitative input from the lecturers regarding the accuracy of extraction and classification of topics, sub-topics, and descriptions through the pipeline. To achieve this quantitative feedback, each lecturer checked a comprehensive list of extracted topics, their sub-topics, and the sub-topic descriptions, and evaluated the following for each sample:
\begin{enumerate}
    \item The correctness of extraction.
    \item The hierarchical allocation of sub-topics within the topic.
\end{enumerate}
Lecturers also evaluated the structural extraction and verified whether the extracted sub-topics were correctly positioned within the corresponding topics.
A total of 1197 extraction samples were evaluated by the experts (173 Topics, 512 Sub-Topics, and 512 Descriptions). Table \ref{Table:1} shows the results of the Precision, Recall, and F1 measure for Topic and Sub-Topic classes in both evaluation modules. For the Description class, we calculate the Precision value to verify whether the long description text that was generated automatically fits the corresponding sub-topic.

The proposed pipeline and LLM-supported extraction of topics and sub-topics has shown a high capability of detecting the topics and their corresponding sub-topics from the lecture materials. A slight difference can be noticed between the two modules, which we trace back to the level and style of structuring that each lecturer uses to organize their lecture recordings. While some recordings only included the lecturer’s explanation of the session’s content, other recordings were simply made of the live, online session, including the discussions with students, their questions, and the lecturer's answers. The difference between these two approaches had a considerable influence on the concept extraction from transcribed videos in the first iteration of LLM extraction and classification. The model extracted concepts from the side discussions, questions, and answers, as well as from the lecturer’s highlights about topics that will be covered in the course. Those topics, however, were not in fact discussed in the corresponding session. To account for this effect, we followed a prompt engineering approach to set explicit rules and limitations for the way each video transcript is handled. The prompt engineering process also defined the ontology classes to the LLM, to clarify the contextual meaning of the Topic and Sub-Topic classes.

It is important to highlight here that the differences among lecturers in the ways they record their lectures are vast, and therefore cannot be only accounted for in the prompt engineering steps. To that end, diversifying the input of the model, e.g., by supporting the video transcriptions with structured lecture slides or references, will be important to ensure a good quality extraction. Moreover, the KG itself can play the role of a supporting data source, e.g., in a retrieval augmented generation (RAG) approach, to extend the context of the LLM’s prompt and increase the relevance and accuracy of the extraction and classification tasks.

\begin{table}[]
\caption{TABLE I. 	PRECISION (P), RECALL (R), AND F1 MEASURE VALUES FOR TOPICS AND SUB-TOPICS OF THE EVALUATION MODULES.}
  \label{Table:1}
\begin{tabular}{|l|lll|lll|l|}
\hline
\multirow{2}{*}{} & \multicolumn{3}{l|}{\textbf{Topic}}                                               & \multicolumn{3}{l|}{\textbf{Sub-Topic}}                                           & \textbf{Description} \\ \cline{2-8} 
                  & \multicolumn{1}{l|}{\textbf{P}} & \multicolumn{1}{l|}{\textbf{R}} & \textbf{F1}   & \multicolumn{1}{l|}{\textbf{P}} & \multicolumn{1}{l|}{\textbf{R}} & \textbf{F1}   & \textbf{P}           \\ \hline
ESa               & \multicolumn{1}{l|}{0,99}       & \multicolumn{1}{l|}{0,94}       & 0,96          & \multicolumn{1}{l|}{1}          & \multicolumn{1}{l|}{0,97}       & \textbf{0,98} & \textbf{1}           \\ \hline
FPGA              & \multicolumn{1}{l|}{0,97}       & \multicolumn{1}{l|}{1}          & \textbf{0,98} & \multicolumn{1}{l|}{0,89}       & \multicolumn{1}{l|}{0,99}       & 0,94          & 0,89                 \\ \hline
\end{tabular}
\end{table}

\subsection{KG-Structural Evaluation}
To evaluate the resulting knowledge graph structure, we use the graph quality measures: Average Degree Centrality (ADC), and Graph Modularity. These measures validate our research goal of discovering connections between learning contents from multiple modules. Average degree centrality evaluates the connectedness of a graph node through its incoming and outgoing relations. In our use case, ADC highlights the role that the semantic similarity algorithm plays in increasing the density of relations among graph nodes. Graph modularity, on the other hand, evaluates the potential to separate learning contents in the graph from each other. Therefore, a lower modularity score is preferred in our case, since it reflects stronger connections between similar contents.

\begin{figure}
  \centering
  \includegraphics[width=0.8\linewidth]{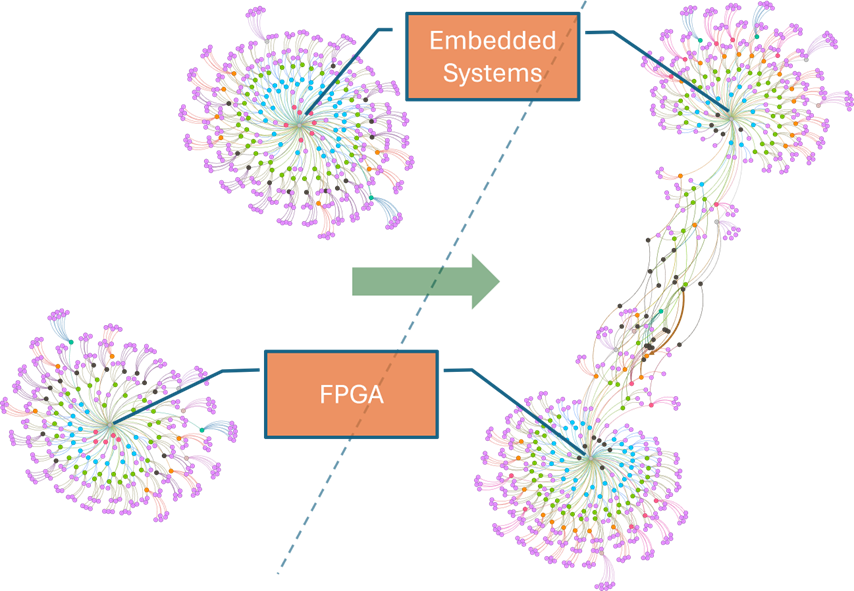}
  \caption{The structure of the KG before (left) and after (right) connecting both evaluation modules through semantic Topic and Sub-Topic relations.}
  \label{Fig:3}
\end{figure}

In the resulting KG, see Figure \ref{Fig:3}, we calculate an increase of the ADC value from 0.9 to 1.03, as well as a slight modularity decrease from 0.769 to 0.767. This slight change can be traced back to the small number of semantic relations added between the two modules, due to the relatively small size of the sample data. Moreover, this is also a result of the high similarity thresholds that we have set for the semantic relation extraction algorithm, to ensure high-quality relations between the topics and sub-topics.

Despite this relatively small improvement on the KG structurally, qualitative expert feedback pointed out that the automatic topic extraction and the possibility to semantically connect topics from different courses is a considerable help for the lecturers in re-structuring their course contents. This is accomplished by being aware of similar content in other courses and modules, which might be overlapping, or can be re-used. The evaluators pointed out that the insights they received from the pipeline’s output did not only support a better structuring of their courses and modules, but also the personalization of the learning experience for individual students, who have different levels of knowledge, scientific backgrounds, and career goals.

\section{Conclusion}
In this paper, we proposed an approach for modelling higher education curriculums and linking them to domain and user models. We use an LLM-assisted pipeline to conduct a KG completion task, which is accomplished through a collaboration between the intelligent model and domain experts. Our goal was to enable a seamless integration of the existing learning content into the curriculum model, to support learning personalization based on the KG.

We implemented our pipeline on a sample of two university modules within embedded systems and FPGA courses. The proposed base ontology, pipeline, and the extraction and classification models were evaluated with the lecturers of the selected evaluation modules. Our evaluation results showed the effectiveness of the proposed approach for the collaboration between human experts and the AI system. It served as an assistant, providing the human expert with an advanced starting point for the structuring of the learning content, and for enabling better personalization of the learning experience, considering the similarities between module contents and their connection to the domain and user models.

\bibliography{sample-1col}

\begin{thebibliography}{13}
\expandafter\ifx\csname natexlab\endcsname\relax\def\natexlab#1{#1}\fi
\providecommand{\url}[1]{\texttt{#1}}
\providecommand{\href}[2]{#2}
\providecommand{\path}[1]{#1}
\providecommand{\DOIprefix}{doi:}
\providecommand{\ArXivprefix}{arXiv:}
\providecommand{\URLprefix}{URL: }
\providecommand{\Pubmedprefix}{pmid:}
\providecommand{\doi}[1]{\href{http://dx.doi.org/#1}{\path{#1}}}
\providecommand{\Pubmed}[1]{\href{pmid:#1}{\path{#1}}}
\providecommand{\bibinfo}[2]{#2}
\ifx\xfnm\relax \def\xfnm[#1]{\unskip,\space#1}\fi
\bibitem[{Meissner and Köbis(2020)}]{bielikova_annotated_2020}
\bibinfo{author}{R.~Meissner}, \bibinfo{author}{L.~Köbis},
\newblock \bibinfo{title}{Annotated {Knowledge} {Graphs} for {Teaching} in {Higher} {Education}: {Supporting} {Mentors} and {Mentees} by {Digital} {Systems}},
\newblock in: \bibinfo{editor}{M.~Bielikova}, \bibinfo{editor}{T.~Mikkonen}, \bibinfo{editor}{C.~Pautasso} (Eds.), \bibinfo{booktitle}{Web {Engineering}}, volume \bibinfo{volume}{12128}, \bibinfo{publisher}{Springer International Publishing}, \bibinfo{address}{Cham}, \bibinfo{year}{2020}, pp. \bibinfo{pages}{551--555}. \URLprefix \url{http://link.springer.com/10.1007/978-3-030-50578-3_43}. \DOIprefix\doi{10.1007/978-3-030-50578-3_43}, \bibinfo{note}{series Title: Lecture Notes in Computer Science}.
\bibitem[{Ibrahim et~al.(2024)Ibrahim, Aboulela, Ibrahim, and Kashef}]{ibrahim_survey_2024}
\bibinfo{author}{N.~Ibrahim}, \bibinfo{author}{S.~Aboulela}, \bibinfo{author}{A.~Ibrahim}, \bibinfo{author}{R.~Kashef},
\newblock \bibinfo{title}{A survey on augmenting knowledge graphs ({KGs}) with large language models ({LLMs}): models, evaluation metrics, benchmarks, and challenges},
\newblock \bibinfo{journal}{Discover Artificial Intelligence} \bibinfo{volume}{4} (\bibinfo{year}{2024}) \bibinfo{pages}{76}. \URLprefix \url{https://link.springer.com/10.1007/s44163-024-00175-8}. \DOIprefix\doi{10.1007/s44163-024-00175-8}.
\bibitem[{Naveed et~al.(2023)Naveed, Khan, Qiu, Saqib, Anwar, Usman, Akhtar, Barnes, and Mian}]{naveed_comprehensive_2023}
\bibinfo{author}{H.~Naveed}, \bibinfo{author}{A.~U. Khan}, \bibinfo{author}{S.~Qiu}, \bibinfo{author}{M.~Saqib}, \bibinfo{author}{S.~Anwar}, \bibinfo{author}{M.~Usman}, \bibinfo{author}{N.~Akhtar}, \bibinfo{author}{N.~Barnes}, \bibinfo{author}{A.~Mian}, \bibinfo{title}{A {Comprehensive} {Overview} of {Large} {Language} {Models}}, \bibinfo{year}{2023}. \URLprefix \url{https://arxiv.org/abs/2307.06435}. \DOIprefix\doi{10.48550/ARXIV.2307.06435}, \bibinfo{note}{version Number: 10}.
\bibitem[{Pan et~al.(2023)Pan, Luo, Wang, Chen, Wang, and Wu}]{pan_unifying_2023}
\bibinfo{author}{S.~Pan}, \bibinfo{author}{L.~Luo}, \bibinfo{author}{Y.~Wang}, \bibinfo{author}{C.~Chen}, \bibinfo{author}{J.~Wang}, \bibinfo{author}{X.~Wu},
\newblock \bibinfo{title}{Unifying {Large} {Language} {Models} and {Knowledge} {Graphs}: {A} {Roadmap}}  (\bibinfo{year}{2023}). \URLprefix \url{https://arxiv.org/abs/2306.08302}. \DOIprefix\doi{10.48550/ARXIV.2306.08302}, \bibinfo{note}{publisher: arXiv Version Number: 3}.
\bibitem[{Abu-Rasheed et~al.(2023)Abu-Rasheed, Dornhöfer, Weber, Kismihók, Buchmann, and Fathi}]{abu-rasheed_building_2023}
\bibinfo{author}{H.~Abu-Rasheed}, \bibinfo{author}{M.~Dornhöfer}, \bibinfo{author}{C.~Weber}, \bibinfo{author}{G.~Kismihók}, \bibinfo{author}{U.~Buchmann}, \bibinfo{author}{M.~Fathi},
\newblock \bibinfo{title}{Building {Contextual} {Knowledge} {Graphs} for {Personalized} {Learning} {Recommendations} {Using} {Text} {Mining} and {Semantic} {Graph} {Completion}},
\newblock in: \bibinfo{booktitle}{2023 {IEEE} {International} {Conference} on {Advanced} {Learning} {Technologies} ({ICALT})}, \bibinfo{publisher}{IEEE}, \bibinfo{address}{Orem, UT, USA}, \bibinfo{year}{2023}, pp. \bibinfo{pages}{36--40}. \URLprefix \url{https://ieeexplore.ieee.org/document/10260850/}. \DOIprefix\doi{10.1109/ICALT58122.2023.00016}.
\bibitem[{Kejriwal(2022)}]{kejriwal_knowledge_2022}
\bibinfo{author}{M.~Kejriwal},
\newblock \bibinfo{title}{Knowledge {Graphs}: {A} {Practical} {Review} of the {Research} {Landscape}},
\newblock \bibinfo{journal}{Information} \bibinfo{volume}{13} (\bibinfo{year}{2022}) \bibinfo{pages}{161}. \URLprefix \url{https://www.mdpi.com/2078-2489/13/4/161}. \DOIprefix\doi{10.3390/info13040161}.
\bibitem[{Kommineni et~al.(2024)Kommineni, König-Ries, and Samuel}]{kommineni_human_2024}
\bibinfo{author}{V.~K. Kommineni}, \bibinfo{author}{B.~König-Ries}, \bibinfo{author}{S.~Samuel}, \bibinfo{title}{From human experts to machines: {An} {LLM} supported approach to ontology and knowledge graph construction}, \bibinfo{year}{2024}. \URLprefix \url{https://arxiv.org/abs/2403.08345}. \DOIprefix\doi{10.48550/ARXIV.2403.08345}, \bibinfo{note}{version Number: 1}.
\bibitem[{Jhajj et~al.(2024)Jhajj, Zhang, Gustafson, Lin, and Lin}]{sifaleras_educational_2024}
\bibinfo{author}{G.~Jhajj}, \bibinfo{author}{X.~Zhang}, \bibinfo{author}{J.~R. Gustafson}, \bibinfo{author}{F.~Lin}, \bibinfo{author}{M.~P.-C. Lin},
\newblock \bibinfo{title}{Educational {Knowledge} {Graph} {Creation} and {Augmentation} via {LLMs}},
\newblock in: \bibinfo{editor}{A.~Sifaleras}, \bibinfo{editor}{F.~Lin} (Eds.), \bibinfo{booktitle}{Generative {Intelligence} and {Intelligent} {Tutoring} {Systems}}, volume \bibinfo{volume}{14799}, \bibinfo{publisher}{Springer Nature Switzerland}, \bibinfo{address}{Cham}, \bibinfo{year}{2024}, pp. \bibinfo{pages}{292--304}. \URLprefix \url{https://link.springer.com/10.1007/978-3-031-63031-6_25}. \DOIprefix\doi{10.1007/978-3-031-63031-6_25}, \bibinfo{note}{series Title: Lecture Notes in Computer Science}.
\bibitem[{Abu-Rasheed et~al.(2024)Abu-Rasheed, Weber, and Fathi}]{abu-rasheed_knowledge_2024}
\bibinfo{author}{H.~Abu-Rasheed}, \bibinfo{author}{C.~Weber}, \bibinfo{author}{M.~Fathi},
\newblock \bibinfo{title}{Knowledge {Graphs} as {Context} {Sources} for {LLM}-{Based} {Explanations} of {Learning} {Recommendations}},
\newblock in: \bibinfo{booktitle}{2024 {IEEE} {Global} {Engineering} {Education} {Conference} ({EDUCON})}, \bibinfo{publisher}{IEEE}, \bibinfo{address}{Kos Island, Greece}, \bibinfo{year}{2024}, pp. \bibinfo{pages}{1--5}. \URLprefix \url{https://ieeexplore.ieee.org/document/10578654/}. \DOIprefix\doi{10.1109/EDUCON60312.2024.10578654}.
\bibitem[{Stancin et~al.(2020)Stancin, Poscic, and Jaksic}]{stancin_ontologies_2020}
\bibinfo{author}{K.~Stancin}, \bibinfo{author}{P.~Poscic}, \bibinfo{author}{D.~Jaksic},
\newblock \bibinfo{title}{Ontologies in education – state of the art},
\newblock \bibinfo{journal}{Education and Information Technologies} \bibinfo{volume}{25} (\bibinfo{year}{2020}) \bibinfo{pages}{5301--5320}. \URLprefix \url{https://link.springer.com/10.1007/s10639-020-10226-z}. \DOIprefix\doi{10.1007/s10639-020-10226-z}.
\bibitem[{Katis et~al.(2018)Katis, Kondylakis, Agathangelos, and Vassilakis}]{gangemi_developing_2018}
\bibinfo{author}{E.~Katis}, \bibinfo{author}{H.~Kondylakis}, \bibinfo{author}{G.~Agathangelos}, \bibinfo{author}{K.~Vassilakis},
\newblock \bibinfo{title}{Developing an {Ontology} for {Curriculum} and {Syllabus}},
\newblock in: \bibinfo{editor}{A.~Gangemi}, \bibinfo{editor}{A.~L. Gentile}, \bibinfo{editor}{A.~G. Nuzzolese}, \bibinfo{editor}{S.~Rudolph}, \bibinfo{editor}{M.~Maleshkova}, \bibinfo{editor}{H.~Paulheim}, \bibinfo{editor}{J.~Z. Pan}, \bibinfo{editor}{M.~Alam} (Eds.), \bibinfo{booktitle}{The {Semantic} {Web}: {ESWC} 2018 {Satellite} {Events}}, volume \bibinfo{volume}{11155}, \bibinfo{publisher}{Springer International Publishing}, \bibinfo{address}{Cham}, \bibinfo{year}{2018}, pp. \bibinfo{pages}{55--59}. \URLprefix \url{https://link.springer.com/10.1007/978-3-319-98192-5_11}. \DOIprefix\doi{10.1007/978-3-319-98192-5_11}, \bibinfo{note}{series Title: Lecture Notes in Computer Science}.
\bibitem[{Ilkou et~al.(2021)Ilkou, Abu-Rasheed, Tavakoli, Hakimov, Kismihók, Auer, and Nejdl}]{hotho_educor_2021}
\bibinfo{author}{E.~Ilkou}, \bibinfo{author}{H.~Abu-Rasheed}, \bibinfo{author}{M.~Tavakoli}, \bibinfo{author}{S.~Hakimov}, \bibinfo{author}{G.~Kismihók}, \bibinfo{author}{S.~Auer}, \bibinfo{author}{W.~Nejdl},
\newblock \bibinfo{title}{{EduCOR}: {An} {Educational} and {Career}-{Oriented} {Recommendation} {Ontology}},
\newblock in: \bibinfo{editor}{A.~Hotho}, \bibinfo{editor}{E.~Blomqvist}, \bibinfo{editor}{S.~Dietze}, \bibinfo{editor}{A.~Fokoue}, \bibinfo{editor}{Y.~Ding}, \bibinfo{editor}{P.~Barnaghi}, \bibinfo{editor}{A.~Haller}, \bibinfo{editor}{M.~Dragoni}, \bibinfo{editor}{H.~Alani} (Eds.), \bibinfo{booktitle}{The {Semantic} {Web} – {ISWC} 2021}, volume \bibinfo{volume}{12922}, \bibinfo{publisher}{Springer International Publishing}, \bibinfo{address}{Cham}, \bibinfo{year}{2021}, pp. \bibinfo{pages}{546--562}. \URLprefix \url{https://link.springer.com/10.1007/978-3-030-88361-4_32}. \DOIprefix\doi{10.1007/978-3-030-88361-4_32}, \bibinfo{note}{series Title: Lecture Notes in Computer Science}.
\bibitem[{Radford et~al.(2022)Radford, Kim, Xu, Brockman, McLeavey, and Sutskever}]{radford_robust_2022}
\bibinfo{author}{A.~Radford}, \bibinfo{author}{J.~W. Kim}, \bibinfo{author}{T.~Xu}, \bibinfo{author}{G.~Brockman}, \bibinfo{author}{C.~McLeavey}, \bibinfo{author}{I.~Sutskever}, \bibinfo{title}{Robust {Speech} {Recognition} via {Large}-{Scale} {Weak} {Supervision}}, \bibinfo{year}{2022}. \URLprefix \url{https://arxiv.org/abs/2212.04356}. \DOIprefix\doi{10.48550/ARXIV.2212.04356}, \bibinfo{note}{version Number: 1}.

\end{thebibliography}

\end{document}